# Interfacial charge accumulation effect on magnetic domain wall nucleation and propagation in a Pt/Co/Pt/Al$_2$O$_3$ structure

Weiwei Lin (林维维),* Nicolas Vernier, Guillaume Agnus, Na Lei (雷娜), Sylvain Eimer and Dafiné Ravelosona[†]

*Institut d'Electronique Fondamentale, Université Paris-Sud - CNRS, Orsay 91405, France*

We report direct observation of charge accumulation effect on magnetization reversal in a Pt/Co(0.5 nm)/Pt(0.5 nm)/Al$_2$O$_3$ structure with perpendicular anisotropy. By imaging magnetic domain with polar Kerr microscopy, we evidence that positive charges accumulating at the Pt/Al$_2$O$_3$ interface result in favoring magnetic domain wall propagation, while negative charges hinder domain wall nucleation and propagation. Our results suggest that magnetic properties in Co layer can be strongly influenced by 5d electron accumulation/depletion in an ultrathin Pt layer.

PACS numbers: 75.60.Jk, 75.70.Cn, 75.85.+t, 85.75.-d

Controlling domain wall dynamics in magnetic wires is currently an important challenge in spintronics. Approaches based on magnetic field [1-6] and spin-polarized current [7-16] have proven very successful in controlling domain wall (DW) nucleation and propagation but they require high power. Recently, efforts have been made to use a voltage to assist the magnetization switching at low power [17-37]. This appealing route involves studies of magnetoelectric coupling in ferroelectric/ferromagnetic structures [17-22], strain in piezoelectric/ferromagnetic structures [23-25] and electric field, i.e. charge transfer in ferromagnetic metal/dielectric structures [26-35]. Voltage control of magnetic DW motion has been reported recently in the ferroelectric/ferromagnetic structures and the piezoelectric/ferromagnetic structures [38-40]. However, little work has focused on electric field assisted DW motion in the ferromagnetic metal/dielectric structure.

By applying a voltage in the ferromagnetic metal/dielectric structure, charge accumulation is induced at the ferromagnetic metal/dielectric interface and an induced electric field decays in the ferromagnetic metal with a typical screening length of about 1 Å. As a result, magnetic properties such as magnetization, Curie temperature, magnetic anisotropy and coercivity can be modified [26-35]. This electric field effect on magnetic properties is large as the ferromagnetic layer is ultrathin since the interface becomes important. Note that this effect results from modifying directly the number of electrons in the ferromagnetic metal.

In this letter, instead of using a pure dielectric layer in direct contact with the ferromagnetic layer, we report the first observation of the effect of charge accumulation in an ultrathin *nonmagnetic* metal on magnetic DW nucleation and propagation in a ferromagnetic/nonmagnetic/dielectric structure. The Co/Pt structure with perpendicular magnetic anisotropy (PMA) is a model system to study the mechanism of magnetization reversal through DW motion and propagation [3,4,6]. The PMA originates from the strong 3d-5d hybridization at the Co/Pt interface [41,42]. It can be expected that the magnetic properties could be influenced by not only the modification of the number of 3d Co electrons but also that of 5d Pt electrons. As shown schematically in Fig. 1, charge accumulation in the ultrathin Pt layer may modify the 5d orbit of Pt electrons and thus, the Co 3d-Pt 5d hybridization can be influenced. In this case, the induced electric field $E$ decays in Pt with a screening length $\lambda$ of about 1 Å and is almost zero in the Co layer.

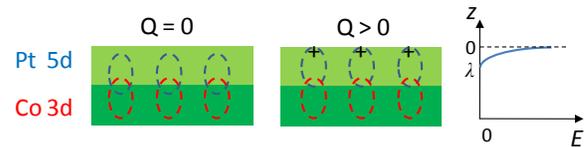

FIG. 1 (color online). Schematic of the 3d-5d hybridization at the Co/Pt interface with and without charge accumulation $Q$ in Pt, and the electric field $E$ induced by the charge accumulation in Pt.

Here, we have focused on imaging magnetic domain motion in a metal/dielectric structure under applied voltage, which is important to provide information of local magnetic domain dynamics, whereas most experiments [26-32] relied on the change in magnetic hysteresis loops under applied voltage. The system studied here is sketched in Fig. 2(a) and 2(b). Firstly, a 350 μm wide Pt(5 nm)/Co(0.5 nm)/Pt(0.5 nm) wire was deposited on top of Si/SiO$_2$ substrate with a shadow mask by magnetron sputtering. Then, a 50 nm thick dielectric Al$_2$O$_3$ film was grown by e-beam evaporation to cover part of the Pt/Co/Pt wire. Finally, a 100 nm thick and 460 μm wide ITO (In$_2$O$_3$:Sn) bar was sputtered on top of Al$_2$O$_3$ and perpendicular to the Pt/Co/Pt wire. The ITO was used as the top electrode because of its relatively good conductivity (about 7 Ω$^{-1}$m$^{-1}$) and transparency, which allows the observation of magnetic domains in the magnetic wire using Kerr microscopy. In spite of the ultrathin Pt layer (0.5 nm), the Pt/Co/Pt wire exhibits a square perpendicular magnetic loop with a typical coercivity of 6 mT at RT, as shown in Fig. 2(c). By

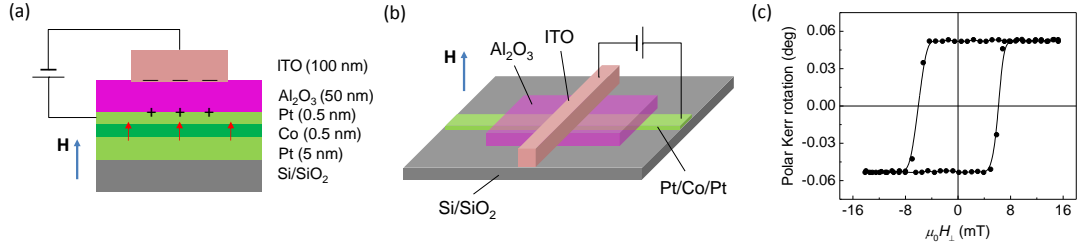

FIG. 2 (color online). (a) and (b) Schematic of the Pt/Co/Pt/Al$_2$O$_3$/ITO structure with applied voltage. (c) Polar Kerr loop of the Pt/Co/Pt film with perpendicular applied magnetic field at room temperature.

applying voltage $V$ between the bottom Pt/Co/Pt and the top ITO, electrons can accumulate or deplete in the ultrathin Pt layer. In our convention, positive (negative) voltage corresponds to positive (negative) charges accumulating at the Pt/Al$_2$O$_3$ interface. The capacitance of the Al$_2$O$_3$ junction is about 0.14 μF/cm$^2$ and the leakage current is less than 8 pA/μm$^2$ for a voltage of 1 V indicating reasonable dielectric properties for the Al$_2$O$_3$ layer. The induced electric field amplitude at the Pt/Al$_2$O$_3$ interface is about 200 kV/cm at 1 V and decays in Pt with $\lambda$ of about 1 Å.

Polar Kerr microscopy was performed at room temprature (RT) with applied magnetic field **H** perpendicular to the film plane, where the positive magnetic field is defined pointing out of the film surface. In our films, magnetization reversal is dominated by domain nucleation followed by easy DW propagation, i.e, the propagation field $H_P$ is smaller than the nucleation field $H_N$. To get a Kerr image of a magnetic domain, a reference image was first taken at zero magnetic field after the sample was saturated at a magnetic field of 20 mT. Then, a magnetic field of −6 mT close to the nucleation field was applied with a duration $\Delta t$ and a second image was taken at zero magnetic field. The final Kerr image was obtained by substracting the two images. The same procedure was repeated under applied voltages.

Figure 3 shows the typical Kerr images of magnetic domains in the cross region of Pt/Co/Pt wire and ITO bar for applied voltages of 0 V, 1 V, and −1 V and for a duration $\Delta t$ of 20 s and 40 s respectively. There are three visible reversed (negative) magnetic domains shown in Fig. 3, where one small domain is in the middle and two large domains move from the edge to the middle. As shown in Fig. 3(a), at $V = 0$ and $H = −6$ mT, a 20 s magnetic field pulse is not sufficient to observe the small nucleated domain, but it is visible after the 40 s magnetic field pulse as shown in Fig. 3(b). This single switching event among a few magnetic domains is then characterized by nucleation of a reversed domain (typical scale of a few tens of nm) followed by DW propagation on a few tens of micrometers. A striking result is that a larger size of the magnetic domain in the middle (indicated by the dash square in Fig. 3) can be observed as a 1 V voltage is applied shown in Fig. 3(c) and 3(d), however for −1 V, the magnetic domain is invisible, as shown in Fig. 3(e) and 3(f). Thus we believe that negative voltage either prevents DW nucleation or DW propagation once a small domain is nucleated. It can be noticed that the magnetic field has an influence on the motion of the two large domains on the sides, whereas the applied voltage has no visible influence. It means that not all magnetic domains in the cross region sample of Pt/Co/Pt wire and ITO bar are influenced by the applied voltage, indicating a non-uniform charge accumulation effect on magnetic properties. This feature may result from the possible inhomogeneous charge density accumulating at the metal/dielectric interface due to the presence of local defects.

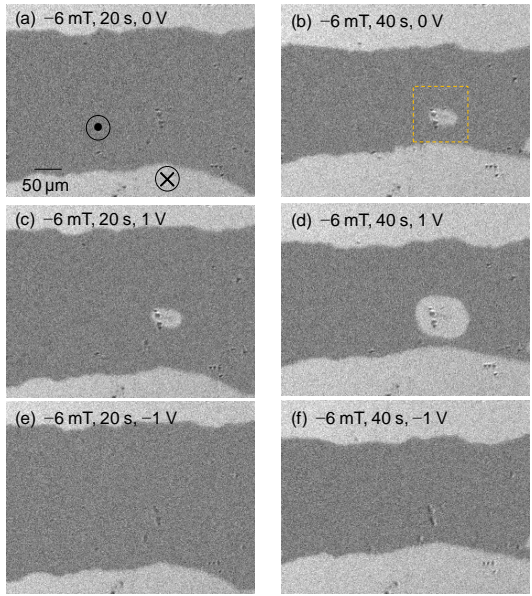

FIG. 3 (color online). Polar Kerr microscopy images of Pt/Co/Pt magnetic domains in the cross region of Pt/Co/Pt and ITO under applied voltages of 0 V, 1 V and −1 V, respectively, for the cases of −6 mT magnetic field pulses with durations of 20 s and 40 s. The dot and cross labels indicate the positive magnetization pointing out of the film surface and the negative one pointing into the film, respectively. The dash square indicates the magnetic domain influenced strongly by the applied voltage.

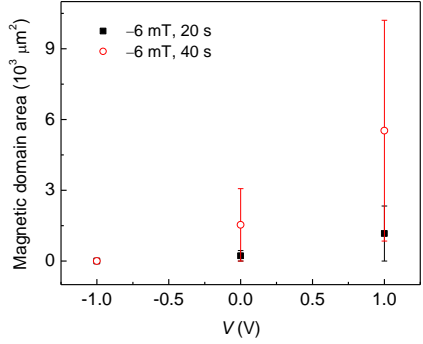

FIG. 4 (color online). Dependence of the area of the magnetic domain (indicated by the dash square in Fig. 3) on the applied voltage in the Pt/Co/Pt/Al$_2$O$_3$/ITO structure, for the cases of −6 mT magnetic field pulses with durations of 20 s (solid squares) and 40 s (open circles), respectively.

By repeating the measurements several times at the same voltage, we observe that magnetization reversal exhibits stochastic behavior with a distribution of the magnetic domain area due to thermal activation. As shown in Fig. 4, for the case of $H = -6$ mT with 40 s duration, the area of the magnetic domain (indicated by the dash square in Fig. 3) ranges from 0 to $3 \times 10^3$ μm$^2$ for $V = 0$, from $1 \times 10^3$ μm$^2$ to $1 \times 10^4$ μm$^2$ for $V = 1$ V, and is always zero for $V = -1$ V. Similar behavior is also observed for the case of −6 mT magnetic field with a shorter pulse duration of 20 s.

These results suggest that positive charges accumulating at the Pt/Al$_2$O$_3$ interface result in magnetic domain wall propagation, while negative charges hinder domain wall nucleation and/or propagation. Since negative and positive applied voltages have clearly an opposite effect here and the leakage current is very low, we can preclude Joule heating influence on magnetization reversal.

To go further in our analysis, the probability of domain wall nucleation and propagation under thermal activation [6] can be expressed as

$$p_N = f_0 \exp\left(-\frac{E_N}{k_B T}\right), \quad p_P = f_0 \exp\left(-\frac{E_P}{k_B T}\right), \quad (1)$$

where $f_0$ is the attempt frequency and $E_N$ ($E_P$) the energy barrier for magnetic DW nucleation (propagation). For Pt/Co/Pt films, the energy barriers can be written as $E_N = 2V_B M_S (H_N - H)$ and $E_P = 2V_B M_S (H_P - H)$ [6], with $V_B$ the Barkhausen volume and $M_S$ the saturation magnetization.

The electron accumulation/depletion in the ultrathin Pt layer can be considered as both charge and spin accumulation and can modify the Pt 5d orbital occupation and spin-dependent density of states. Such effect is expected to influence the spin and orbital moment as well as the Co 3d-Pt 5d hybridization at the Co/Pt interface, so that the magnetization and magnetic anisotropy of the thin Co layer can be modified. The saturation magnetization of $M_S$ of Pt/Co/Pt is about $1.8 \times 10^6$ A/m. The spin moment of bulk Co is about 1.64 $\mu_B$/atom, the orbital moment of fcc Co is about 0.11 $\mu_B$/atom, and the Pt moment is about 0.2 $\mu_B$/atom [42]. Recent experiment [32] shows that electric field at the Co/MgO interface can control the ferromagnetic phase transition in Pt/Co. Particularly, the positive charge in Co layer reduces the Curie temperature of Co by 10 K as applying an electric field of about 2 MV/cm (a change of 0.01 electrons per Co atom). In our case, assuming that the charge accumulation in Pt is uniform, we calculate an average change of about 0.002 electrons per Pt atom (charge density is about 0.14 μC/cm$^2$) at $V = 1$ V. Therefore, we expect the change in $M_S$ due to charge accumulation to be negligible. As for the effect of the electric field on magnetic anisotropy, a recent theoritical work [34] shows that the magnetocrystalline anisotropy of Pt/Co/Pt structure decreases with increasing the number of positive charges in the Pt layer. Since the propagation field and the nucleation field are strongly related to the magnetic anisotropy, our results are consistent with a reduction (increase) of the magnetic anisotropy under a positive (negative) voltage.

Our experiment suggests that the positive charge at the Pt/Al$_2$O$_3$ interface reduces $E_P$, while the negative charge increases $E_P$. Assuming that electron accumulation doesn't affect $V_B$, the reduction of $E_P$ by the positive charge in Pt may result mainly from the reduction of $H_P$, which is related to the local pinning potential of the defect and depends on magnetic anisotropy [6]. Considering $S^V$ and $S^0$ are the magnetic domain areas with and without applied voltage, respectively, from Eq. (1) we can write

$$\ln\frac{S^V}{S^0} = \frac{E_P^0 - E_P^V}{k_B T} = \frac{2V_B M_S}{k_B T}\left(H_P^0 - H_P^V\right) \quad (2)$$

From our results, at $V = 1$ V, $S^V/S^0 \sim 3.3$ and thus, $E_P^0 - E_P^V \sim 1.2 k_B T$. $V_B$ can be estimated about $10^{-18}$ cm$^3$. This suggests that the reduction of propagation field $H_P^0 - H_P^V$ is about 1.6 mT at $V = 1$ V. For the magnetic domain indicated by the dash square in Fig. 3, the domain wall velocity can be estimated roughly about 0.7 μm/s at a magnetic field value of 6 mT under 0 V, and increases up to 1.2 μm/s under 1 V, whose enhancement is about 70%.

In conclusion, we have shown that the energy barrier for magnetic domain wall nucleation and propagation can be strongly influenced by electron accumulation/depletion in an ultrathin nonmagnetic metallic layer. This suggests that the modification of 5d electrons occupation in an ultrathin Pt layer may influence the magnetic properties of Co due to 3d-5d hybridization. Our results also indicate that charge accumulation effect may be inhomogeneous over the film structure, so interpretation of measurements focusing on change of coercivity measured through hysteresis loop has to be taken very carefully. Local measurement of single switching event is a prerequisite to fully understand the electric field effect in magnetic metals. Finally, our results suggest that electric field effect can be used to assist magnetization reversal in domain wall based non-volatile memories or logic devices.

This work was partly supported by French National Research Agency (ANR) ELECMADE project and the European FP7 program through contract MAGWIRE number 257707.

*weiwei.lin@u-psud.fr

†dafine.ravelosona@u-psud.fr